# Past, Present and Future Stars that can see Earth as a Transiting Exoplanet

L. Kaltenegger[1,2]*, J. K. Faherty[3]

[1]Carl Sagan Institute, Cornell University, Space Science Institute 302, 14850 Ithaca, NY, USA,
[2]Astronomy Department, Cornell University, Space Science Institute 302, 14850 Ithaca, NY, USA
[3]Department of Astrophysics, American Museum of Natural History, Central Park West 79th st., NY 10024, USA
*email: lkaltenegger@astro.cornell.edu

**In the search for life in the cosmos, transiting exoplanets are currently our best targets. In the search for life in the cosmos, transiting exoplanets are currently our best targets. With thousands already detected, our search is entering a new era of discovery with upcoming large telescopes that will look for signs of life in the atmospheres of transiting worlds. However, the universe is dynamic, and which stars in the solar neighborhood have a vantage point to see Earth as a transiting planet[1–4] and can identify its vibrant biosphere since early human civilizations are unknown.**

**Here we show that 1,715 stars within 326 light-years are in the right position to have spotted life on a transiting Earth since early human civilization, with an additional 319 stars entering this special vantage point in the next 5,000 years. Among the stars are 7 known exoplanet hosts that hold the vantage point to see Earth transit, including Ross-128, which saw Earth transit in the past, Teegarden's Star, and Trappist-1, which will start to see Earth transit in 29 and 1,642 years, respectively. We found that human-made radio waves have swept over 75 of the closest stars on our list already.**

To obtain the dynamic picture of which stars can see the Earth transit[1–4] and how many years such a viewpoint holds, we evaluated the kinematics propagated through time using ESA's Gaia Mission early Data Release 3 (eDR3)[5].

We identify 2,034 (https://github.com/jfaherty17/ETZ) stars in the Gaia Catalog of Nearby Stars (GCNS)[6] that are in the Earth transit zone (ETZ)[3] over +/- 5,000years: 313 objects were in the ETZ in the past, 319 will enter the ETZ in the future, and 1,402 have been in the ETZ for some time. Given that Earth began transmitting radio waves into the solar neighborhood about 100years ago[7], we also identify the sub-sample of 75 stars located in a 30pc sphere that Earth's radio signals have washed over.

We find a large diversity among the 2,034 objects that enter/exit the ETZ over +/5,000years. Using the Gaia colors and absolute magnitude, $M_G$[8,9] (Figure 1), we estimate stellar spectral types and find that M dwarfs dominate our sample. That agrees with our understanding of the initial mass-distribution[10] for the Milky Way disk population. The 2,034 objects contain 194 G stars like our Sun and 12A, 2B, 87F, 102K, 1,520M stars, 7L, 1T, and 109 white dwarfs (WD) (see Table 1). At least 12 stars are also on the giant branch. There are numerous well-studied stars in this list (e.g., Wolf 359, Ross 128, Teegarden's star) as well as previously unknown nearby sources identified in Gaia eDR3 (e.g., Gaia EDR3 4116504399886241792).

1,402 sources can currently see Earth transit, including 128 G stars like our Sun, 10A, 1B, 63F, 73K, 1050M stars, 2L, and 75WDs. SETI observations of stars in the ETZ have recently been started[11,12].

Isotopic data indicate that life on Earth started by about 3.8 to 3.5 Gya[13,14], which coincides with the end of the heavy bombardment phase[15]. While we cannot infer the time needed for life to start on any exoplanet from Earth's history, the early signs for life on Earth are encouraging. Our ETZ catalog includes all potentially habitable world host stars regardless of mass or evolved state given the uncertainty of how/when/where life might arise and survive[16–20].

Of the 2,034 objects in the ETZ, 67% (1,355) are also in the restricted ETZ (rETZ)[3] - which guarantees a minimum view of Earth's transit for 10 hours (see Table 1): 311 were in the rETZ in the past, 330 will enter the rETZ in the future, and 713 have been in the rETZ in the +/-5,000-year time window. This sample is mainly composed of cool stars, with colors and magnitudes consistent with 6A, 57F, 121G, 72K, 1,021M stars, 5L, 1T, and 72WDs.

Out of the 2,034 objects in the ETZ, 117 objects lie within 30pc (~ 100 light-years) of the Sun. Among those sources, 29 were in the ETZ in the past, 42 will enter it in the future, and 46 have been in the ETZ for some time. These 46 objects would be able to see Earth transit while also being able to detect emitted radio waves from Earth[7], which would have reached those stars by now: 2F, 3G, 2K, 34M stars, and 5WDs (see Table 1).

Seven of the 2,034 stars are known exoplanet host stars: table 2 shows them sorted by distance from the Sun. Four of the planet-hosts are located within 30pc (~100 light-years).

The Ross128 system at 3.375pc is the 13$^{th}$ closest system to the Sun and the second closest system with a transiting Earth-size exoplanet. Ross128 would have seen Earth transit for 2,158years – from 3,057 until 900years ago. Teegarden's star system, the 25$^{th}$ closest system to the Sun at 3.832pc, hosts two non-transiting Earth-mass worlds. It will enter the ETZ in 29years[21] for 410years. The GJ9066 system at 4.470pc hosts one non-transiting planet, will enter the ETZ in 846years and remain in it for 932years. The Trappist-1 system at 12.467pc hosts 7 Earth-size planets, with 4 of them in the temperate, habitable zone (HZ). This system will enter the ETZ in 1,642years and remain for 2,371years.

Three more exoplanet systems beyond 30pc but within 100pc have been in the ETZ for thousands of years: K2-240 at 73.043pc hosts two transiting planets. The system entered the ETZ more than 5,000years ago and will remain in the ETZ well past 5,000years into the future. K2-65 at 63.104pc, with one transiting planet, entered the ETZ 2,183years ago and will stay in the ETZ past 5,000years in the future. K2-155 at 72.932pc hosts three transiting planets. The system entered the ETZ more than 5,000years ago and will remain for another 3,118years.

Estimates of the number of rocky planets in the HZ[22] in the literature are given for various values of a planet's radius limit and HZ limits[23,24]. New estimates place it from 0.58+0.73/-0.33 to 0.88+1.28/-0.51 planets per star[24] for the empirical HZ. Then, the Empirical HZ limits are set by the flux a young Mars and a young Venus received when there is no more evidence for liquid water on their surfaces[22]. The inner limit is not well known because of the lack of reliable geological surface history for Venus beyond 1 billion years ago.

While the discussion on the occurrence rate of rocky planets is ongoing, here, we use a pessimistic rate of 25% to estimate that there are potentially 508 rocky worlds in the HZ of our full sample. Restricting the selection to the distance radio waves from Earth have traveled- about 100 light-years - leads to an estimated 29 potentially habitable worlds that could have seen Earth transit and also detect radio waves from our planet.

We found that 43% (868) of stars in our sample spend at least 10,000years, 68% (1,380) at least 5,000years, and 94% (1,910) at least 1,000years in the ETZ. Sources closer to the Sun have larger proper motion values than those further away, thus stars in our closest neighborhood move in and out of the ETZ faster than more distant objects. The average

ETZ viewing window for the 100pc sample over the 10,000years evaluated is 6,914years, while the average for just the 30pc sample is 3,973years. However, our analysis only provides a first assessment of the time a star keeps its vantage point to observe Earth as a transiting planet. Only 78 objects complete an entrance and exit over the +/-5000-year period.

The universe is dynamic, changing the vantage point of other stars to find us over thousands of years – as well as ours to find planets transiting other stars.

During Earth's Anthropocene period – 237 years so far, starting with the steam engine in 1784[25,26], when humans started to influence Earth's climate – 1,424 stars have seen the Earth transit. For a further 1,000 and 5,000-year period, that number would increase to 1,489 and 1,743, respectively.

The discussion on whether or not we should send out an active signal or try to hide our presence is ongoing[27,28]. However, our biosphere has modified our planet's atmosphere for billions of years[13,14], something we hope to find on other Earth-like planets soon. Thus, observing Earth as a transiting planet could have classified it as a living world since the great oxidation event for a billion years already[29–31], through the buildup of oxygen in the presence of a reducing gas[23,32–36]. Here we assume that any nominal civilization on an exoplanet would have astronomical instrumentation comparable to what we have now.

Recent technosignatures[37] - like radio waves – can, in addition, indicate a technological civilization on Earth. Even though humans have emitted radio waves only for a comparably short time - about 100years - they have already reached 75 stars in the past and present ETZ in our neighborhood, traveling further out still.

Our analysis shows that even the closest stars generally spend more than 1,000years at a vantage point where they can see Earth transit - and thus, we can assume the reverse will also be true. That provides a long timeline for nominal civilizations to identify Earth as an interesting planet.

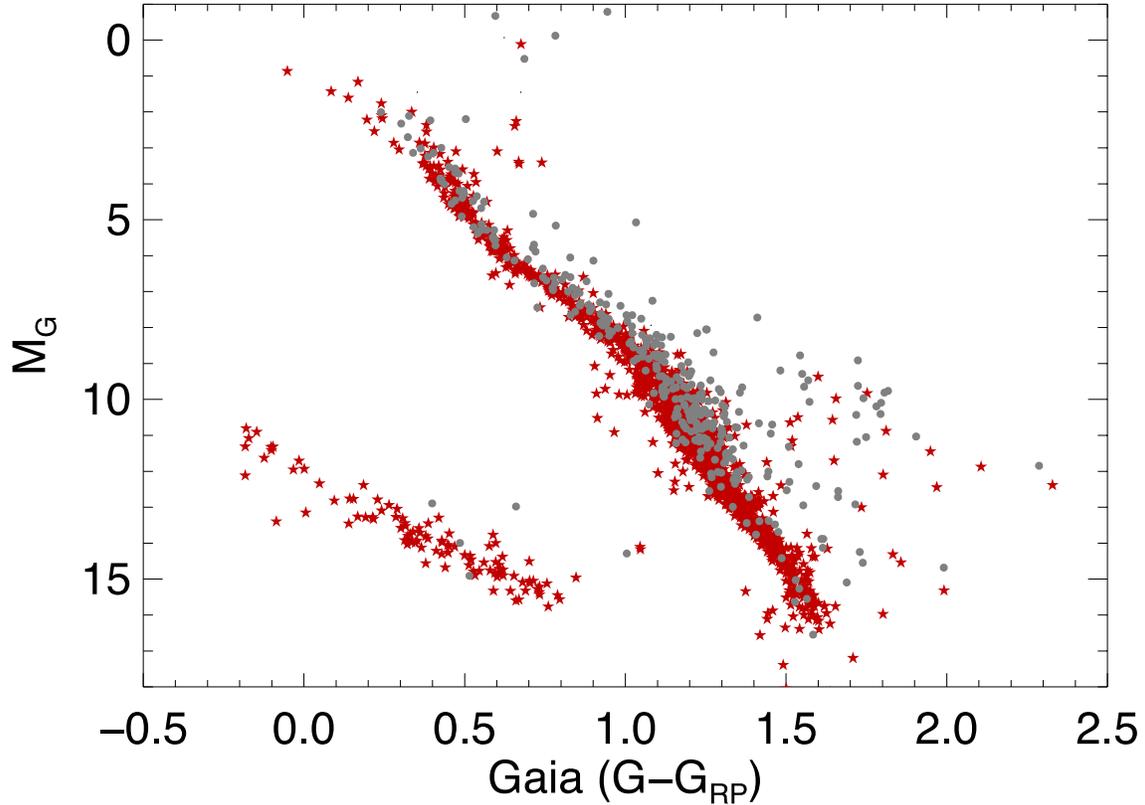

**Figure 1: Stars that can see Earth transit since early human civilization**: The color-magnitude diagram of the Gaia Catalogue of Nearby Stars (black) limited to sources with RUWE<1.4, photometric signal to noise (in Gaia G and $G_{RP}$)>100, and parallax uncertainties better than 5%. Overplotted are the 2,034 sources that cross the ETZ in the time interval of +/-5,000years (dark red five-point stars have RUWE<1.4, grey filled circles RUWE>1.4). $M_G$ is Gaia magnitude. Data at: https://github.com/jfaherty17/ETZ.

**Table 1: Sample of the full machine-readable table of ETZ stars sorted by distance** [1]

| Star Name | | Distance | Earth Transit Zone (ETZ) | | | |
|---|---|---|---|---|---|---|
| **Name** | **GAIA eDR3** | **[lightyears]** | **Enter [yr]** | **Exit [yr]** | **Total [yr]** | **When** |
| Wolf 359 | 3864972938605115520 | 7.9 | -46 | 432 | 480 | Past to future |
| Ross 128 | 3796072592206250624 | 11 | -3057 | -900 | 2158 | Past |
| Teegardens star | 35227046884571776 | 12.5 | 29 | 438 | 410 | Future |

**Table 2: Sample of the full machine-readable table of exoplanet hosts in the ETZ sorted by distance**

| Star Name | | Distance | Earth Transit Zone (ETZ) | | | |
|---|---|---|---|---|---|---|
| **Exoplanet Host** | **GAIA eDR3** | **[light years]** | **Enter [yr]** | **Exit [yr]** | **Total [yr]** | **When** |
| Ross 128 | 3796072592206250624 | 11 | -3057 | -900 | 2158 | Past |
| Teegarden's Star | 35227046884571776 | 12.5 | 29 | 438 | 410 | Future |
| GJ 9066 | 76868614540049408 846 | 14.6 | 846 | 1777 | 932 | Future |
| TRAPPIST-1 | 2635476908753563008 | 40.6 | 1642 | 4012 | 2371 | Future |
| K2-65 | 2613211076737129856 | 205.8 | -2183 | 5000 | 7184 | Past to future |
| K2-155 | 145333927996558976 | 237.9 | -5000 | 3118 | 8119 | Past to future |
| K2-240 | 6257625719430982016 | 238.2 | -5000 | 5000 | 10000 | Past to future |

---

[1] full table in Supplement Material and at https://github.com/jfaherty17/ETZ

**Methods**

The European Space Agency's Gaia mission has revolutionized our understanding of the local solar neighborhood. Gaia released over 1.3 billion sources with 5-parameter astrometric solutions in data release 2 (proper motion and parallax values)[8]. The Gaia catalog's early DR3 in Dec 2020[5] contained the Gaia Catalogue of Nearby Stars (GCNS)[6]. The GCNS is a self-defined "clean" and well-characterized collection of objects within 100pc of the Sun, including at least 92% of stars of stellar types down to M9[6].

We converted right ascension and declination values to ecliptic latitude and longitude values for the total 100 pc GCNS sample to determine which stars could see Earth transit from their position. The Earth transit zone (ETZ)[3] corresponds to a thin strip around the ecliptic as projected onto the sky with an ecliptic latitude width of 0.528 degrees. The smaller restricted Earth Transit zone (rETZ)[3] with an ecliptic latitude width of 0.264 degrees denotes the region from which the Earth's transit can be seen for a minimum of 10 hours.

In the 302,197 stars in the 100pc GCNS sample, we identify 1,402 objects currently located at ecliptic latitude between +/-0.264 degrees. We compared our sample with an earlier study[4] that found 1,004 main sequence stars in the ETZ using the Transiting Exoplanet Survey Satellite (TESS)[38] Input Catalog (TIC)[39] matched to Gaia DR2[8]. We find that in the updated eDR3 Gaia catalog, 21 of the original objects are now either outside of 100 pc (20) or are unreported in Gaia eDR3 (1). Thus, our sample has 983 sources in common with the earlier sample of stars that can see Earth transit now. Note that the earlier 1004 stars sample[4] excluded evolved stars from their selection and limited the Gaia quality flag in the TIC to unity.

The reliability of the astrometry is critical to determining the robustness of a candidate in the ETZ. As such, we use the re-normalized unit weight error (RUWE) value from the catalog to flag potentially poor candidates. As recommended[6], a RUWE near unity is the signature of a robust astrometric solution. Values larger than 1.4 are considered suspect for various reasons, including unresolved binarity, variability, and crowding.

Among our full sample of 2,034 objects from eDR3 (https://github.com/jfaherty17/ETZ), which are in the ETZ for the +/-5000-year period, 349 objects have a RUWE >1.4. Like the procedure used to discover the nearby L dwarf WISE J192512.78+070038.8[40] which was in the Galactic plane, we visually inspected each source using the citizen scientist developed image viewing tool called Wiseview[41] to ensure that crowding did not insert a false detection.

We also examined the Gaia color-magnitude diagram (CMD) for all targets to flag any that appeared in suspect areas. Figure 1 shows the CMD for the GCNS catalog (black) - for (G-$G_{RP}$) colors – limited to sources with RUWE<1.4, photometric signal to noise (in Gaia G and $G_{RP}$)>100, and parallax uncertainties better than 5%. Overplotted are the 2,034 sources that cross the ETZ in the time interval of +/-5,000years (dark red five-point stars have RUWE<1.4, grey filled circles have RUWE>1.4). Data are available at https://github.com/jfaherty17/ETZ. We also examined the Gaia ($G_{BP}$-$G_{RP}$) CMD to complement the analysis. Both Gaia CMD's show scatter across the stars, which is a function of several parameters such as varying stellar temperatures, binarity, and age[9,42,43]. Note that the (G-$G_{RP}$) color is discerning for the lowest temperature sources (e.g., M dwarfs), while the ($G_{BP}$-$G_{RP}$) color is helpful for a close examination of the higher mass sources. We primarily used the (G-$G_{RP}$) CMD to determine a spectral type (SpT) estimate of each source; however, we turned to the ($G_{BP}$-$G_{RP}$) color for confirmation in the case of higher mass stars and white dwarfs.

The vast majority of our ETZ sources lie within the full GCNS sample's scatter on the CMD. However, a small number of sources show photometry or astrometry that is likely suspect even for objects with RUWE<1.4, given the significant number of objects with Gaia $(G-G_{RP})> 1.5$. That might be a sign of a given object's intrinsic properties (like age, metallicity, binarity). Table 1 lists parameters for the full 2,034 ETZ sample (data at https://github.com/jfaherty17/ETZ), including relevant Gaia astrometric and photometric information and our estimation of the SpT for each source from its CMD position. We also cross-matched our full sample against literature estimates of mass, effective temperature, radii, bolometric luminosity, metallicity, and log g for Gaia sources. Table 1 lists all values with respective catalog references for the parameter noted[39,44–46].

To identify stars that have seen and will see Earth as a transiting planet, we propagate motions backward and forward in time. To do this, we used the RA, DEC, parallax, $\mu_{ra}$, and $\mu_{dec}$ values from the GCNS. Note that we did not eliminate sources that were CMD or RUWE outliers (see Table 1 for detailed information). We used the full 100pc sample and iterated in 1-year bins backward and forward in time. We list all astrometric information, including RUWE, in Table 1 so the reader can decide on a criterion of their choosing for follow-up of the sample.

Ideally, to conduct this analysis, we would use full spatial and velocity information (XYZUVW). However, only a small subsample of objects in Gaia eDR3 have radial velocity measurements. Therefore, we had to proceed with only tangential velocities and the projected positions across the sky over time for this work. While we cannot account for sources moving toward or away from us, in a short time frame analysis, the tangential motion of a source across the sky drives the impact on its position in the ETZ. Therefore, radial velocity has a minimal effect on the analysis. However, to truly understand the volume of sources analyzed, one would want all astrometric components. A dedicated RV campaign for these objects will be necessary to accomplish that.

To compile a list of past and future host stars in the ETZ and how long they spent with a view of Earth transiting, we converted the GCNS positions into XYZ cartesian coordinates using subroutines from the Banyan Sigma kinematic analysis code[47]. We then propagated the XYZ positions forward and backward using proper motion values from Gaia eDR3 and a zero value for the radial velocity. We then converted XYZ cartesian coordinates back to ecliptic longitude and latitude in intervals of one year and checked whether the host stars had entered/exited the ETZ. We used the visualization tool called OpenSpace[48] to examine the sample as we proceeded with iterations. Using that software we could move time forward and backward and see how the sample changed in position over time. In this way we could visually confirm all stars. Using this iterative method, we found 2,034 potential exoplanet host stars in the ETZ in the past, present, and future.

Given that all stars are in motion around the center of the Galaxy, over time, the linear projection of a star's motion will deviate significantly from the circular orbit a star is taking around the Galaxy. The solar system's estimated time to complete one orbit is ~250Myr. Using a geometric approximation we find that after ~150Myr, the deviation between the linear approximation and the circular orbit would approach the size of the ETZ window. Therefore, we chose a conservative +/-5000-years period for our analysis. This timeframe both conservatively accommodates the projected positions over time and includes a critical portion of human development on Earth. Furthermore, +/-5,000years ensures that we cover ample time for another nominal civilization to have detected and studied Earth through the rise of modern human civilization.

We approximated the ETZ and rETZ entrance and exit windows' sizes using the uncertainties in astrometric parameters. For each of the 2034 stars in our sample, we ran 100 iterations of their position propagated using randomly sampled one sigma uncertainties in RA, DEC, parallax, $\mu_{ra}$, and $\mu_{dec}$. We evaluated the ETZ and rETZ entrance and exit points for each of the 100 iterations and reported the standard deviation of the sample as the window uncertainties in Table 1. The majority of stars have such small astrometric errors that the windows have negligible uncertainties. Moreover, with 43% of objects spending more than 10,000years in the ETZ, many stars enter and exit the ETZ well before/after our analysis.

The length of time that any given star within 100pc stays in the ETZ in our analysis is proper motion, which generally scales with distance. Thus, sources closer to the Sun have generally larger proper motion values than those further away. For instance, the median total proper motion for the 30pc sample in the GCNS is about 300 mas yr$^{-1,}$ whereas the median value for objects out to 100pc is about 85 mas yr$^{-1.}$ That means that sources closer to the Sun will move through the ETZ faster than those at larger distances. We find that over these 10,000years, only 78 objects complete an entrance and exit. Most stars in our sample (1,954) have either entered the ETZ before our analysis started or entered the ETZ and will stay beyond our analysis timeframe.

109 of the objects in our catalog are White dwarfs, dead stellar remnants. While most searches for life on other planets concentrate on main sequence stars[23,34,35], the recent discovery of a giant planet around a white dwarf[19] opened the intriguing possibility that we might also find rocky planets orbiting evolved stars[16–20,49]. Characterizing rocky planets in the Habitable Zone of a WD would answer intriguing questions on lifespans of biota or a second "genesis" after a star's death[50].

Stars with a vantage point that could see Earth transit - and thus see an interesting planet for deliberate broadcasts - are priority targets for searches for life and extraterrestrial intelligence[1–4]. First observations of such stars have recently been started[11,12]. Also, NASA's TESS has entered the extended mission phase, with a plan to observe stars across the ecliptic, including those with a vantage point to see Earth transit.

**References Methods**

**Acknowledgments**
L.K. acknowledges support from the Carl Sagan Institute at Cornell and the Breakthrough Initiative. J.K.F. acknowledges support from the Heising Simons Foundation and the Research Corporation for Science Advancement (Award 2019-1488). This work has made use of data from the European Space Agency (ESA) mission Gaia processed by the Gaia Data Processing and Analysis Consortium DPAC20 https://www.cosmos.esa.int/gaiahttps://www.cosmos.esa.int/web/gaia/dpac/consortium Funding for the DPAC has been provided by national institutions, in particular the institutions participating in the Gaia Multilateral Agreement. This research has also used NASA's Astrophysics Data System and the VizieR and SIMBAD databases operated at CDS, Strasbourg, France.

**Author contributions**
L.K. envisioned the idea of the paper, J.K.F identified the ETZ stars. L.K. and J.K.F composed the manuscript, undertook the analysis, and discussed the content of this manuscript.

**Data and code availability**
All data is available in the supplement material. In addition, all data and code can be found on https://github.com/jfaherty17/ETZ.

**Competing Interests**
The authors declare that they have no competing financial interests.

**Additional Information**
Supplementary Information is available for this paper.

**Stars that can see Earth transit in the +/-5000-year period.** Table 1 lists all characteristics for stars that can see Earth transit in the +/-5000-year period, sorted by distance from the Sun. We also cross-matched our full sample against literature estimates of mass, effective temperature, radii, bolometric luminosity, metallicity, and log g for Gaia sources. Respective catalog references[6,39,44-46] for the parameter are noted in Table 1.

**Exoplanet host stars that can see Earth transit in the +/-5000-year period.** Table 2 lists the characteristics of the seven known stars that can see Earth transit in the +/-5000-year period, that are known exoplanet host stars, sorted by distance from the Sun.

Correspondence and requests for materials should be addressed to L.K. (email: lkaltenegger@astro.cornell.edu).

Reprints and permissions information is available at www.nature.com/reprints